\documentclass[aps]{revtex4}
\usepackage{amsfonts}
\usepackage{amsmath}
\usepackage{amssymb,epsf}

\begin{document}

\title{Extended phase space thermodynamics and $P-V$ criticality\\
of charged black holes in Brans-Dicke theory}
\author{S. H. Hendi$^{1,2}$\footnote{email address: hendi@shirazu.ac.ir} and Z. Armanfard$^{1}$}
\affiliation{$^1$ Physics Department and Biruni Observatory,
College of Sciences, Shiraz University, Shiraz 71454, Iran\\
$^{2}$ Research Institute for Astronomy and Astrophysics of
Maragha (RIAAM), Maragha, Iran}

\begin{abstract}
In this paper, taking in to account Brans-Dick theory, we
investigate thermodynamic behavior of charged black hole
solutions. We study the analogy of the black hole solution with
the Van der Waals liquid-gas system in the extended phase space by
considering the cosmological constant as dynamical pressure. We
obtain critical values of thermodynamic coordinates and plot
$P-r_{+}$ and $G-T$ diagrams to study the phase transition.
\end{abstract}

\maketitle

\section{Introduction\label{Intro}}

Einstein constructed the theory of general relativity that
describes the dynamics of our solar system well enough, but it
probably does not describe gravity accurately at all scales. The
problem that general relativity faced is that, it does not
accommodate either Mach's principle or Dirac's large-number
hypothesis. It is unable to describe the accelerated expansion of
the universe accurately. Herein cosmologists explored various
alternative theories of gravity \cite{MOG}. Brans and Dicke were
pioneers in studying these alternative theories and they developed
another relativistic theory known as Brans-Dicke (BD) theory
\cite{BD}. This theory can be regarded as an economic modification
of Einstein general relativity which describes gravitation in
terms of metric as well as a scalar field and it accommodate both
Mach's principle and Dirac's large- number hypothesis. Due to the
importance of black holes and gravitational collapse in both
classical and quantum gravity, authors have investigated various
aspects of them in BD theory \cite{Gao3}. It has been proved that
in four dimensions, the stationary and vacuum BD solution is just
the Kerr solution with a constant scalar field \cite{Gao4}. In
order to investigate the distinction between the BD theory and
Einstein theory Cai and Myung proved that the black hole solution
in the BD--Maxwell theory in four dimensions is just the
Reissner--Nordstr\"{o}m (RN) solution with a trivial scalar field
\cite{sh5}. In higher dimensions, however, it would be the RN
solution with a non-trivial scalar field. This is because the
stress energy tensor of Maxwell field is not traceless in higher
dimensions and the action of Maxwell field is not invariant under
the conformal transition.

On the other hand, thermodynamic properties of the black holes
have been fascinating subject for many years. It was found out
that black holes along all assigned thermodynamic variables also
have rich phase structure in complete analogy with
non-gravitational thermodynamic system similar to van der Waals
gas system. With the conception of expecting the cosmological
constant term to arise from the vacuum expectation value of a
quantum field, we can assume that it can vary. Hence, we can treat
the cosmological constant and its conjugate as dynamical pressure
and volume of a black hole system respectively
\cite{hendivahid1,hendivahid2,hendivahid3}. Studying the
thermodynamics of black holes in AdS space time has exhibit
various phase transitions with the same critical behavior as van
der Waals model, qualitatively \cite{hendivahid6}. The paper of
Hawking was the initial studies on this subject \cite{kubiznak1}.
He pointed out there is a thermal radiation (black hole first
order phase transition) for Schwarzschild- AdS black hole space
time. Adding charge and/or radiation
will result a behavior similar to a van der Waals liquid/gas \cite%
{kubiznak2,kubiznak3,kubiznak5} and the analogy will improve by
being in the extended phase space where the cosmological constant
is interpreted as thermodynamical pressure.

In this paper, we want to investigate the thermodynamic phase
transition of charged black holes in BD theory by using the
analogy between our system and the van der Waals liquid/gas.

The outline of our paper is as follows. Section \ref{FE} is
devoted to brief review of BD--Maxwell field equations with their
relation with dilaton gravity by a conformal transformation. In
Sec. \ref{Sol}, we obtain charged black hole solutions of both
dilaton gravity and BD theory. Next, we extend the phase space by
considering cosmological constant as thermodynamic pressure and
calculate critical values, and then we plot diagrams for different
cases in Sec. \ref{P-V}. In next section, we give a detailed
discussion regarding diagrams, their physical interpretations, and
the effects of BD parameter. We finish our paper with some closing
remarks.

\section{Field equations and conformal transformations \label{FE}}

The action of $(n+1)$- dimensional BD--Maxwell theory with a
scalar field $\Phi$ and a self-interacting potential $V(\Phi)$ can
be written as \cite{sh5}
\begin{equation}
I_{G}=-\frac{1}{16\pi }\int_{\mathcal{M}}d^{n+1}x\sqrt{-g}\left( \Phi
\mathcal{R}\text{ }-\frac{\omega }{\Phi }(\nabla \Phi )^{2}-V(\Phi )-F_{\mu
\nu }F^{\mu \nu }\right) ,  \label{acBD}
\end{equation}
where $\mathcal{R}$ is the scalar curvature, the factor $\omega $
is the coupling constant, $F_{\mu \nu }=\partial _{\mu }A_{\nu
}-\partial _{\nu }A_{\mu }$ is the electromagnetic tensor field
and $A_{\mu }$ is the electromagnetic potential. Varying the
action (\ref{acBD})
with respect to the gravitational field $g_{\mu \nu }$, the scalar field $%
\Phi $ and the gauge field $A_{\mu }$, one can obtain equations of
motion with the following explicit forms \cite{sh5}
\begin{eqnarray}
G_{\mu \nu } &=&\frac{\omega }{\Phi ^{2}}\left( \nabla _{\mu }\Phi \nabla
_{\nu }\Phi -\frac{1}{2}g_{\mu \nu }(\nabla \Phi )^{2}\right) -\frac{V(\Phi )%
}{2\Phi }g_{\mu \nu }+\frac{1}{\Phi }\left( \nabla _{\mu }\nabla _{\nu }\Phi
-g_{\mu \nu }\nabla ^{2}\Phi \right)  \nonumber \\
&&+\frac{2}{\Phi }\left( F_{\mu \lambda }F_{\nu }^{\text{ }\lambda }-\frac{1%
}{4}F_{\rho \sigma }F^{\rho \sigma }g_{\mu \nu }\right) ,  \label{field01}
\end{eqnarray}
\begin{equation}
\nabla ^{2}\Phi =-\frac{n-3}{2\left[ \left( n-1\right) \omega +n\right] }%
F^{2}+\frac{1}{2\left[ \left( n-1\right) \omega +n)\right] }\left[ (n-1)\Phi
\frac{dV(\Phi )}{d\Phi }-\left( n+1\right) V(\Phi )\right] ,  \label{field02}
\end{equation}
\begin{equation}
\nabla _{\mu }F^{\mu \nu }=0,  \label{field03}
\end{equation}
where $G_{\mu \nu }$ and $\nabla _{\mu }$ are, respectively, the Einstein
tensor and covariant derivative of manifold $\mathcal{M}$ with metric $%
g_{\mu \nu}$. Due to the appearance of the second derivatives of scalar
field in the right hand side of (\ref{field01}), solving the field equations
(\ref{field01})-(\ref{field03}) directly is a non-trivial task. Using a
suitable conformal transformation, one can remove this difficulty. Indeed,
via the conformal transformation \cite{sh5} the BD--Maxwell theory can be
transformed into the Einstein- Maxwell theory with a minimally coupled
scalar dilaton field. Suitable conformal transformation can be shown as
\begin{eqnarray}
\bar{g}_{\mu \nu } &=&\Phi ^{2/(n-1)}g_{\mu \nu },  \nonumber \\
\bar{\Phi} &=&\frac{n-3}{4\alpha }\ln \Phi ,  \label{con}
\end{eqnarray}
where
\begin{equation}
\alpha =(n-3)/\sqrt{4(n-1)\omega +4n}.  \label{a}
\end{equation}

It is notable that all functions and quantities in Jordan frame (${g%
}_{\mu \nu }$, ${\Phi}$ and ${F}_{\mu \nu }$) can be transformed into
Einstein frame ($\bar{g}_{\mu \nu }$, $\bar{\Phi}$ and $\bar{F}_{\mu \nu }$%
). Applying the mentioned conformal transformation on the BD action (\ref%
{acBD}), one can obtain action of dilaton gravity
\begin{equation}
\bar{I}_{G}=-\frac{1}{16\pi }\int_{\mathcal{M}}d^{n+1}x\sqrt{-\bar{g}}%
\left\{ \bar{\mathcal{R}}-\frac{4}{n-1}(\bar{\nabla }\bar{\Phi})^{2}-\bar{V}(%
\bar{\Phi})-\exp \left( -\frac{4\alpha \bar{\Phi}}{(n-1)}\right) \bar{F}%
_{\mu \nu }\bar{F}^{\mu \nu }\right\} ,  \label{con-ac}
\end{equation}
where $\bar{\mathcal{R}}$ and $\bar{\nabla}$ are, respectively, the Ricci
scalar and covariant derivative corresponding to the metric $\bar{g}_{\mu
\nu}$, and $\bar{V}(\bar{\Phi})$ is
\begin{equation}
\bar{V}(\bar{\Phi})=\Phi^{-(n+1)/(n-1)}V(\Phi).  \label{poten}
\end{equation}

Regarding $(n+1)-$dimensional Einstein--Maxwell--dilaton action (\ref{con-ac}%
), $\alpha$ is an arbitrary constant that governs the strength between the
dilaton and Maxwell fields. One can obtain the equations of motion by
varying this action (\ref{con-ac}) with respect to $\bar{g}_{\mu \nu }$, $%
\bar{\Phi}$ and $\bar{F}_{\mu \nu}$
\begin{equation}
\bar{\mathcal{R}}_{\mu \nu }=\frac{4}{n-1}\left( \bar{\nabla}_{\mu }\bar{\Phi%
}\bar{\nabla}_{\nu }\bar{\Phi}+\frac{1}{4}\bar{V}\bar{g}_{\mu \nu }\right)
+2e^{-4\alpha \bar{\Phi}/(n-1)}\left( \bar{F}_{\mu \lambda }\bar{F}_{\nu }^{%
\text{ }\lambda }-\frac{1}{2(n-1)}\bar{F}_{\rho \sigma }\bar{F}^{\rho \sigma
}\bar{g}_{\mu \nu }\right) ,  \label{fieldc1}
\end{equation}
\begin{equation}
\bar{\nabla}^{2}\bar{\Phi}=\frac{n-1}{8}\frac{\partial \bar{V}}{\partial
\bar{\Phi}}-\frac{\alpha }{2}e^{-4\alpha \bar{\Phi}/(n-1)}\bar{F}_{\rho
\sigma }\bar{F}^{\rho \sigma },  \label{fieldc2}
\end{equation}
\begin{equation}
\partial _{\mu }\left[ \sqrt{-\bar{g}}e^{-4\alpha \bar{\Phi}/(n-1)}\bar{F}
^{\mu \nu }\right] =0  \label{fieldc3}
\end{equation}

By assuming the $\left(\bar{g}_{\mu \nu },\bar{F}_{\mu \nu },\bar{\Phi }
\right) $ as solutions of Eqs. (\ref{fieldc1})-(\ref{fieldc3}) with
potential $\bar{V}\left( \bar{\Phi}\right)$ and comparing Eqs. (\ref{field01}%
)-(\ref{field03}) with Eqs. (\ref{fieldc1})-(\ref{fieldc3}) we find the
solutions of Eqs. (\ref{field01})-(\ref{field03}) with potential $V(\Phi)$
can be written as
\begin{equation}
\left[ g_{\mu \nu },F_{\mu \nu },\Phi \right] =\left[ \exp \left( -\frac{%
8\alpha \bar{\Phi}}{\left( n-1\right) (n-3)}\right) \bar{g}_{\mu \nu },\bar{F%
}_{\mu \nu },\exp \left( \frac{4\alpha \bar{\Phi}}{n-3}\right)
\right]. \label{sol}
\end{equation}

\section{Charged solutions in $\left( n+1\right) $-- dimensions\label{Sol}}

Our strategy is to construct the solutions of BD theory with
$n\geqslant 4$ and the quadratic potential
\begin{equation}
V(\Phi )=2\Lambda \Phi^{2}.
\end{equation}

Applying the conformal transformation (\ref{con}), the potential $\bar{V}(%
\bar{\Phi})$ becomes a Liouville-type potential
\begin{equation}
\bar{V}(\bar{\Phi})=2\Lambda \exp \left( \frac{4\alpha \bar{\Phi}}{n-1}%
\right) .  \label{liovilpoten}
\end{equation}

In other words, instead of solving Eqs. (\ref{field01})-(\ref{field03}) with
quadratic potential, we solve Eqs. (\ref{fieldc1})-(\ref{fieldc3}) with
Liouville-type potential. Assuming the $(n+1)-$dimensional metric
\begin{equation}
d\bar{s}^{2}=-f(r)dt^{2}+\frac{dr^{2}}{f(r)}+r^{2}R^{2}(r)d\Omega _{n-1}^{2},
\label{metric}
\end{equation}
where $d\Omega _{n-1}^{2}$ is the metric of a unit $(n-1)-$sphere, and $f(r)$
and $R(r)$ are metric functions. By integrating the Maxwell equation (%
\ref{fieldc3}), we can obtain the nonzero electric field
$\bar{F}_{tr}$ as
\begin{equation}
\bar{F}_{tr}=\frac{q}{(rR)^{n-1}}\exp \left( \frac{4\alpha \bar{\Phi}}{n-1}%
\right).  \label{maxwell}
\end{equation}

Taking into account the metric (\ref{metric}) with Maxwell field (\ref%
{maxwell}), the solutions of (\ref{fieldc1}) and (\ref{fieldc2})
are
\begin{eqnarray}
f(r) &=&-\frac{\left( n-2\right) \left( \alpha ^{2}+1\right) ^{2}c^{-2\gamma
}r^{2\gamma }}{\left( \alpha ^{2}+n-2\right) \left( \alpha ^{2}-1\right) }+%
\frac{2\Lambda (\alpha ^{2}+1)^{2}c^{2\gamma }}{(n-1)(\alpha ^{2}-n)}%
r^{2(1-\gamma )}-\frac{m}{r^{(n-2)}}r^{(n-1)\gamma }  \nonumber \\
&&+\frac{2q^{2}(\alpha ^{2}+1)^{2}c^{-2(n-2)\gamma }}{(n-1)(\alpha
^{2}+n-2)r^{2(n-2)(1-\gamma )}},  \label{F(r)} \\
R(r) &=&\exp (\frac{2\alpha \bar{\Phi}}{n-1})=\left( \frac{c}{r}\right)
^{\gamma },  \label{R(r)} \\
\bar{\Phi}(r) &=&\frac{(n-1)\alpha }{2(1+\alpha ^{2})}\ln (\frac{c}{r}),
\label{phi}
\end{eqnarray}
where $m$ is an integration constant which is related to the total mass, $c$
is another arbitrary constant related to the scalar field and $\gamma=\alpha
^{2}/(1+\alpha ^{2})$.

Now, we are in a position to obtain the solutions of Eqs. (\ref{field01})-(%
\ref{field03}) by using the conformal transformation. Considering the
following spherically symmetric metric
\begin{equation}
ds^{2}=-U(r)dt^{2}+\frac{dr^{2}}{V(r)}+r^{2}H^{2}(r)d\Omega _{n-1}^{2},
\label{metric1}
\end{equation}
with Eqs. (\ref{field01})-(\ref{field03}), we find that the
functions $U(r)$ and $V(r)$ are
\begin{eqnarray}
U(r) &=&\frac{2\Lambda (\alpha ^{2}+1)^{2}c^{2\gamma (\frac{n-5}{n-3})}}{%
(n-1)(\alpha ^{2}-n)}r^{2(1-\frac{\gamma \left( n-5\right) }{n-3})}-\frac{%
mc^{(\frac{-4\gamma }{n-3})}}{r^{(n-2)}}r^{\gamma (n-1+\frac{4}{n-3})}
\nonumber \\
&&+\frac{2q^{2}(\alpha ^{2}+1)^{2}c^{-2\gamma (n-2+\frac{2}{n-3})}}{%
(n-1)(\alpha ^{2}+n-2)r^{2[(n-2)(1-\gamma )-\frac{2\gamma }{n-3}]}}-\frac{%
\left( n-2\right) \left( \alpha ^{2}+1\right) ^{2}}{\left( \alpha
^{2}+n-2\right) \left( \alpha ^{2}-1\right) }\left( \frac{c}{r}\right)
^{-2\gamma \left( \frac{n-1}{n-3}\right) },  \label{u(r)} \\
V(r) &=&\frac{2\Lambda (\alpha ^{2}+1)^{2}c^{2\gamma (\frac{n-1}{n-3})}}{%
(n-1)(\alpha ^{2}-n)}r^{2(1-\frac{\gamma \left( n-1\right) }{n-3})}-\frac{%
mc^{(\frac{4\gamma }{n-3})}}{r^{(n-2)}}r^{\gamma
(n-1-\frac{4}{n-3})}  \nonumber
\\
&&+\frac{2q^{2}(\alpha ^{2}+1)^{2}c^{-2\gamma (n-2-\frac{2}{n-3})}}{%
(n-1)(\alpha ^{2}+n-2)r^{2[(n-2)(1-\gamma )+\frac{2\gamma }{n-3}]}}-\frac{%
2\left( n-2\right) \left( \alpha ^{2}+1\right) ^{2}}{\left( \alpha
^{2}+n-2\right) \left( \alpha ^{2}-1\right) }\left(
\frac{c}{r}\right) ^{-2\gamma \left( \frac{n-5}{n-3}\right) }.
\label{V(r)}
\end{eqnarray}

Using the conformal transformation the electromagnetic field
becomes
\begin{equation}
F_{tr}=\frac{qc^{(3-n)\gamma }}{r^{(n-3)(1-\gamma )+2}}.  \label{ftr}
\end{equation}

As one can see electromagnetic field becomes zero as
$r\longrightarrow \infty $. It is also notable that obtained
solutions are just the charged solutions of Einstein gravity
(Reissner- Nordstr\"{o}m AdS black hole) as $\omega
\longrightarrow \infty $.

\section{Extended phase space and P- V criticality in BD black holes \label{P-V}}

Calculations show that the Hawking temperature of a BD black hole
on the outer horizon $r_{+}$ is
\begin{equation}
T=\frac{\kappa }{2\pi }=\left. \frac{1}{4\pi }\sqrt{\frac{V}{U}}\left( \frac{%
dU}{dr}\right) \right\vert _{r=r_{+}}, \label{tp}
\end{equation}
where $\kappa$ is the surface gravity. After some simplifications,
we obtain
\begin{eqnarray}
T &=&-\frac{2(1+\alpha ^{2})}{4\pi (n-1)}\left( \Lambda b^{2\gamma
}r_{+}^{1-2\gamma }+\frac{q^{2}b^{-2(n-2)\gamma }}{{r_{+}}^{\gamma }}{r_{+}}%
^{(2n-3)(\gamma -1)}\right)  \nonumber \\
&&+\frac{\left[ \gamma \left( n-3\right) -n+2\right] \left( \alpha
^{2}+1\right) ^{2}\left( n-2\right) }{4\pi r_{+}\left( \alpha
^{2}+n-2\right) \left( \alpha ^{2}-1\right) }\left( \frac{c}{r_{+}}\right)
^{-2\gamma },  \label{temp1}
\end{eqnarray}
which is invariant under the conformal transformation because the
conformal parameter is regular at the horizon. The finite mass and
the entropy of the black hole can be obtained by using the
Euclidian action \cite{sh5}
\begin{eqnarray}
M&=&\frac{c^{(n-1)\gamma }}{16\pi }\left( \frac{n-1}{1+\alpha
^{2}}\right) m,   \label{m} \\
S&=&\frac{c^{(n-1)\gamma }}{4}r_{+}^{(n-1)\left( 1-\gamma
\right)}. \label{s}
\end{eqnarray}

We regard $\Lambda $ and its corresponding conjugate quantity as
the thermodynamic pressure $P=\frac{-\Lambda }{8\pi }$ and volume
respectively
\begin{equation}
V=\left( \frac{\partial H}{\partial P}\right) _{S,Q}=\left( \frac{\partial M%
}{\partial P}\right)_{S,Q}. \label{V1}
\end{equation}

Herein, we are interested in studying the phase transition of this
black hole. The equation of state of the black hole can be obtain
using equation (\ref{temp1})
\begin{eqnarray}
P &=&\frac{q^{2}\left( n-1\right) }{8\pi r_{+}^{\left( n-1\right) }}\left(
\frac{b}{r_{+}}\right) ^{-2\gamma \left( n-1\right) }+\frac{T\left(
n-1\right) }{4\left( \alpha ^{2}+1\right) r_{+}}\left( \frac{b}{r_{+}}%
\right) ^{-2\gamma }  \nonumber \\
&&-\frac{\left( n-1\right) \left( n-2\right) \left( \alpha ^{2}+1\right) %
\left[ \gamma \left( n-3\right) -n+2\right] }{16\pi \left( \alpha
^{2}-1\right) \left( \alpha ^{2}+n-2\right) }r_{+}^{2\left( \gamma -1\right)
}\left( \frac{bc}{r_{+}}\right) ^{-2\gamma },  \label{P}
\end{eqnarray}
where $r_{+}$ is linear function of the specific volume $v$ in
geometric unit \cite{hendivahid6}.

We can investigate the existence of phase transition and critical
behavior of this black hole by plotting and analyzing the graphs
of $P-v$ and $G-T$ diagrams. One may use the inflection point
properties
\begin{eqnarray}
&&\left( \frac{\partial P}{\partial v}\right) _{T}=0,  \nonumber \\
&&\left( \frac{\partial ^{2}P}{\partial v^{2}}\right) _{T}=0,
\nonumber
\end{eqnarray}
to obtain the critical values for the temperature, pressure and
volume. Due to the difficulties of solving these equations
analytically, we use the numerical method to obtain critical
values.

According to first law of black hole thermodynamics and the
interpretation of $M$ (total mass of black hole) as $H$ (the black
hole enthalpy) \cite{Gibbs} the Gibbs free energy of black hole
can be written as
\begin{equation}
G=H-TS=M-TS
\end{equation}
\begin{eqnarray}
G &=&\frac{\left( \alpha ^{2}+1\right) (n-1)\left( n-2\right) r_{+}^{n-2}}{%
16\pi \left( \alpha ^{2}+n-2\right) \left( 1-\alpha ^{2}\right) }\left(
\frac{c}{r_{+}}\right) ^{\gamma \left( n-3\right) }\left( 1+\frac{\left(
\alpha ^{2}+1\right) \left[ \left( \gamma -1\right) n-3\gamma +2\right] }{n-1%
}\right)  \nonumber \\
&&-\frac{P\left( \alpha ^{2}+1\right) }{r_{+}^{\gamma \left( n+1\right) -n}}%
\left( \frac{b^{2\gamma }c^{\gamma \left( n-1\right) }}{n-1}+\frac{c^{\gamma
\left( n+1\right) }}{\alpha ^{2}-n}\right)  \nonumber \\
&&-\frac{q^{2}r_{+}^{\gamma \left( n-3\right) -n+2}}{\pi }\left( \frac{%
c^{\gamma \left( n-1\right) }b^{-2\gamma \left( n-2\right)
}}{\left( n-1\right) }+\frac{c^{-\gamma \left( n-3\right)
}}{\left( \alpha ^{2}+n-2\right) }\right).  \label{G}
\end{eqnarray}

The behavior of Gibbs free energy with respect to temperature may
be investigated by plotting the graph of $G-T$. We will see the
characteristics swallow-tail behavior which guarantees the
existence of the phase transitions.

\begin{figure}[tbp]
$%
\begin{array}{ccc}
\epsfxsize=6cm \epsffile{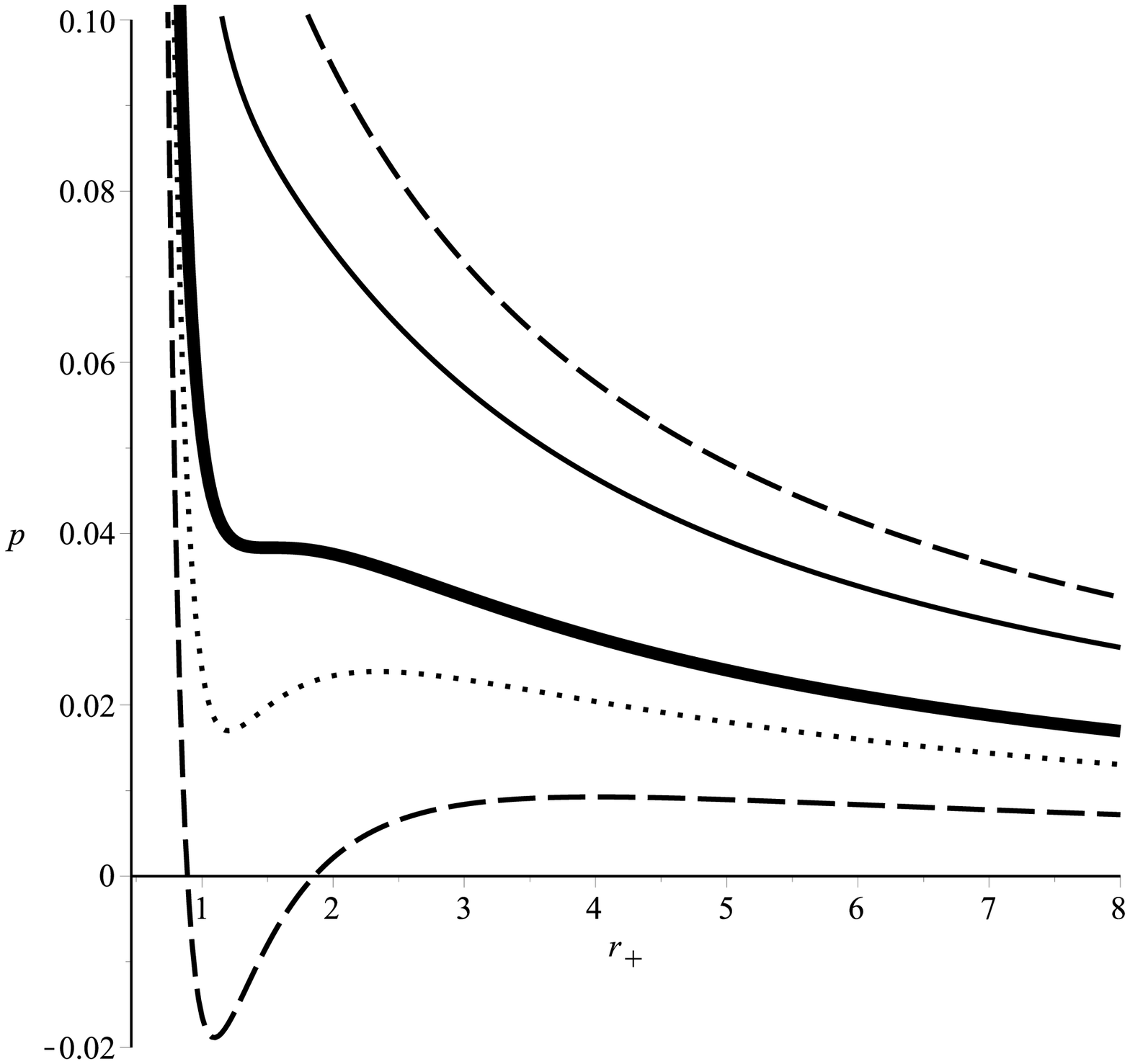} & \epsfxsize=6cm \epsffile{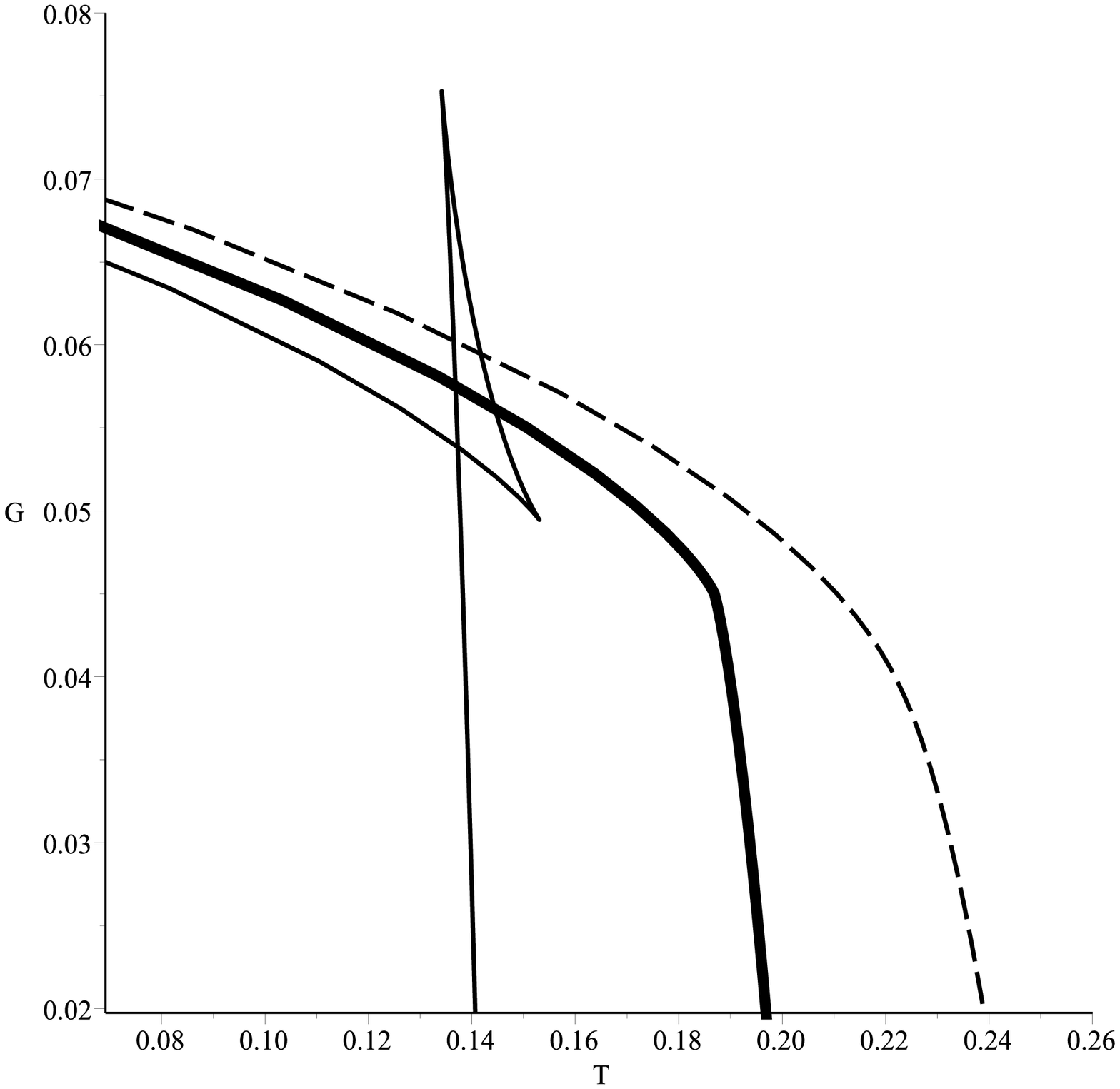}
&
\end{array}
$%
\caption{$P-r_{+}$ (left), $G-T$ (right) diagrams for $c=1$,
$n=4$, $q=1$ and $\protect\omega =1$.\newline
$P-r_{+}$ diagram, from up to bottom $T=1.8T_{c}$, $T=1.5T_{c}$, $T=T_{c}$, $%
T=0.8T_{c}$ and $T=0.5T_{c}$ respectively.\newline $G-T$ diagram,
from up to bottom $P=1.5P_{c}$, $P=P_{c}$ and $P=0.5P_{c}$,
respectively.} \label{Figw1n4}
\end{figure}
\begin{figure}[tbp]
$%
\begin{array}{ccc}
\epsfxsize=6cm \epsffile{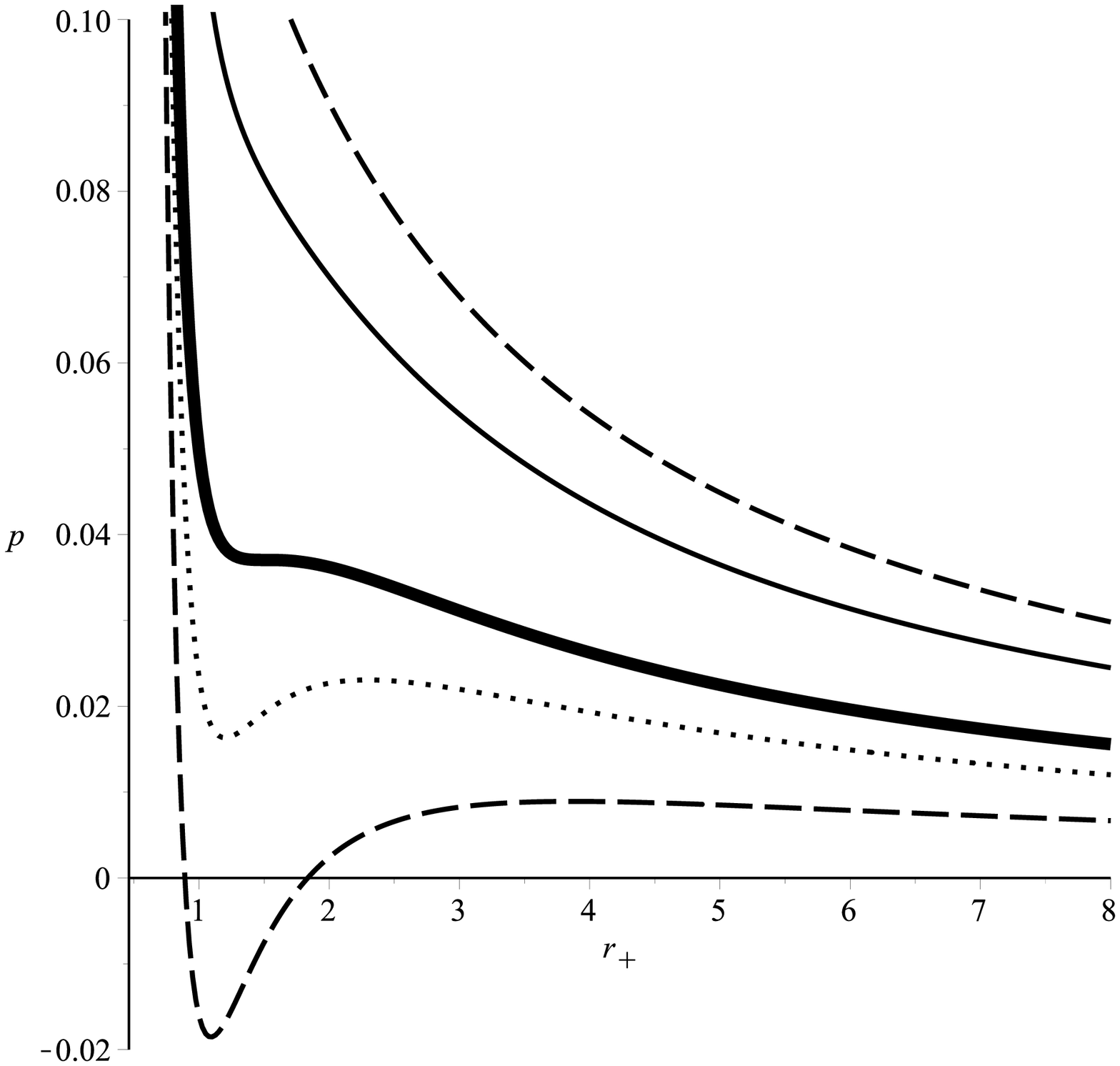} & \epsfxsize=6cm \epsffile{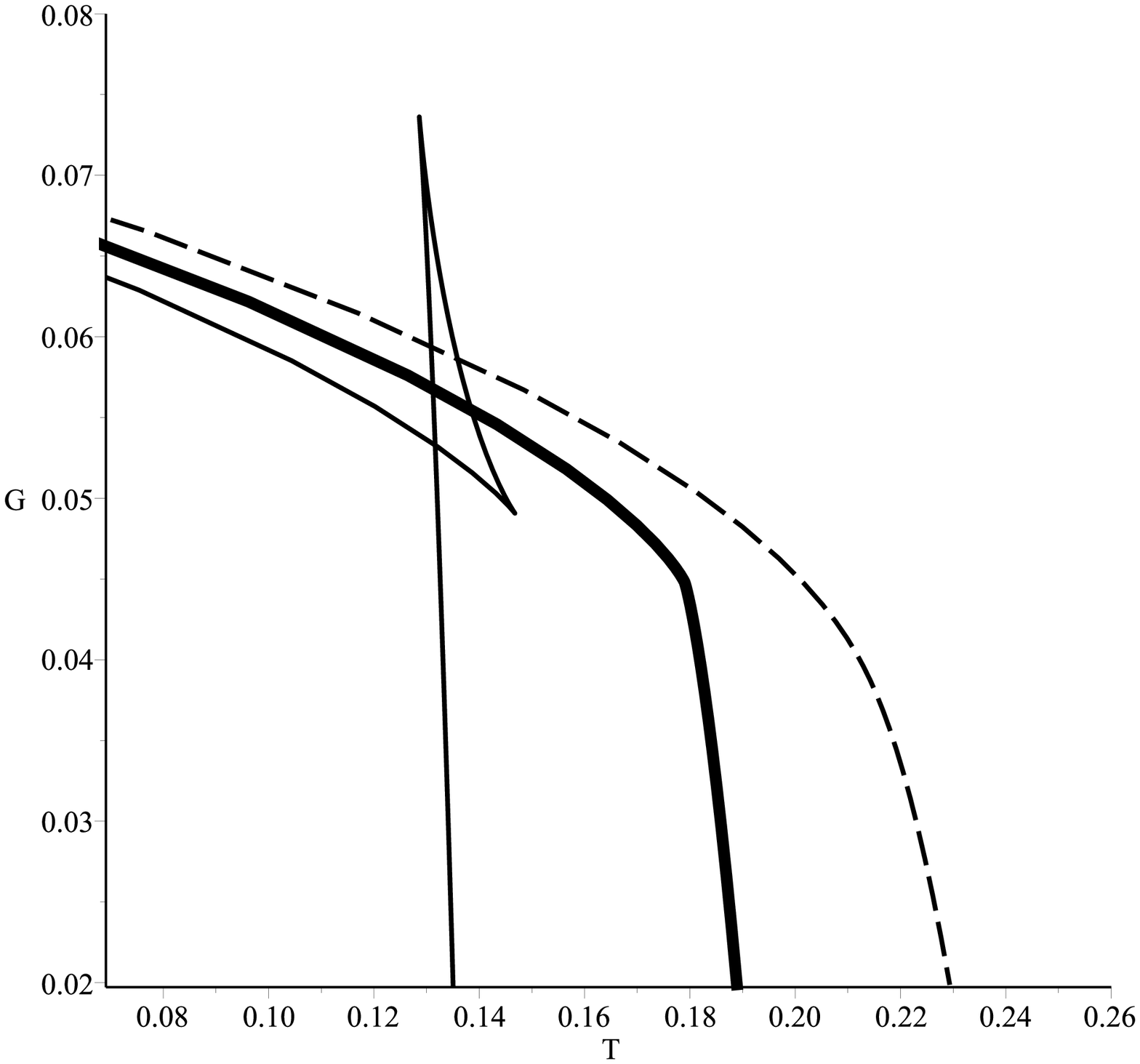}
&
\end{array}
$%
\caption{$P-r_{+}$ (left), $G-T$ (right) diagrams for $c=1$, $n=4$, $q=1$ and $%
\protect\omega =3$.\newline
$P-r_{+}$ diagram, from up to bottom $T=1.8T_{c}$, $T=1.5T_{c}$, $T=T_{c}$, $%
T=0.8T_{c}$ and $T=0.5T_{c}$ respectively.\newline $G-T$ diagram,
from up to bottom $P=1.5P_{c}$, $P=P_{c}$ and $P=0.5P_{c}$,
respectively.} \label{Figw3n4}
\end{figure}
\begin{figure}[tbp]
$%
\begin{array}{ccc}
\epsfxsize=6cm \epsffile{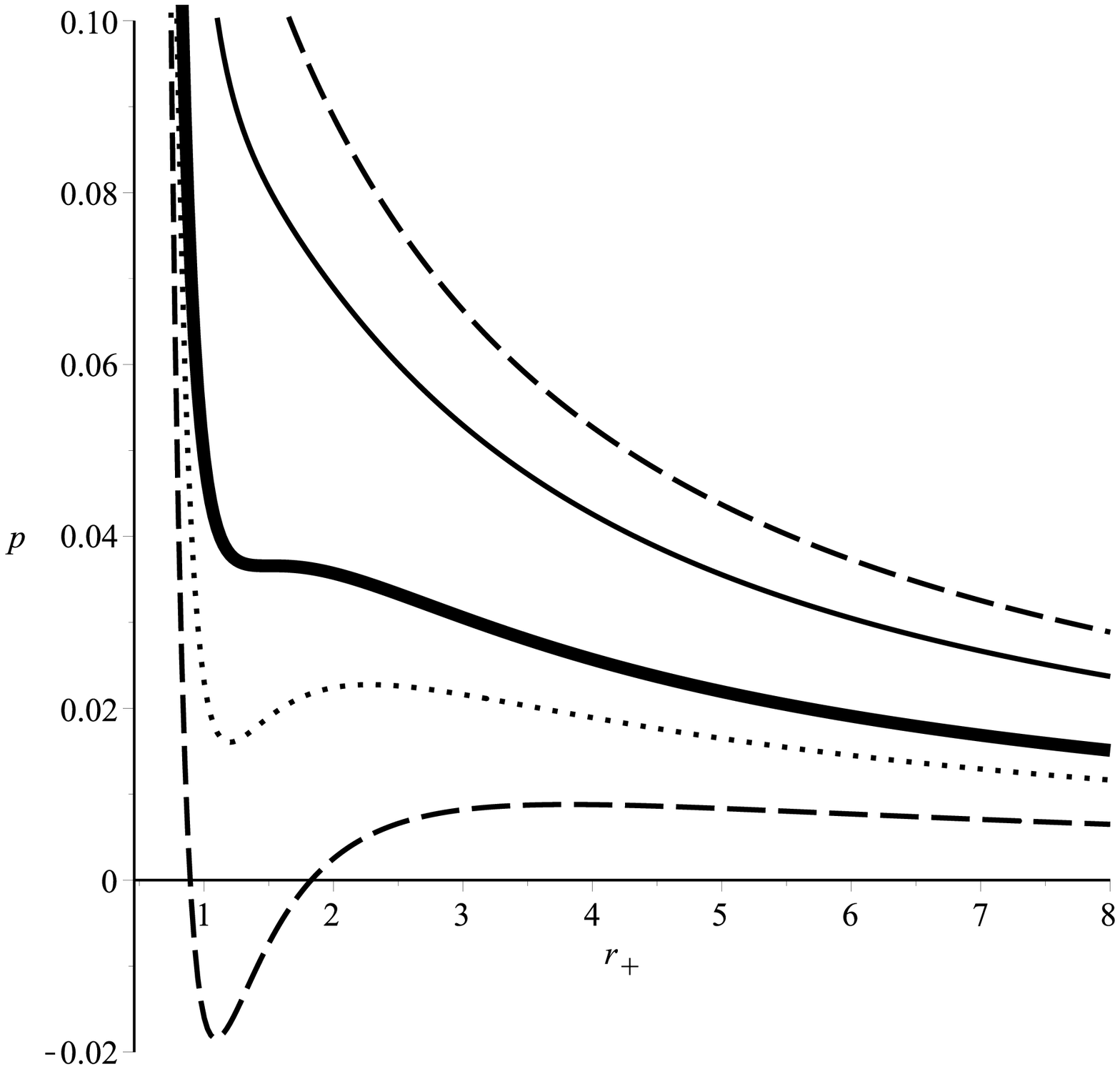} & \epsfxsize=6cm \epsffile{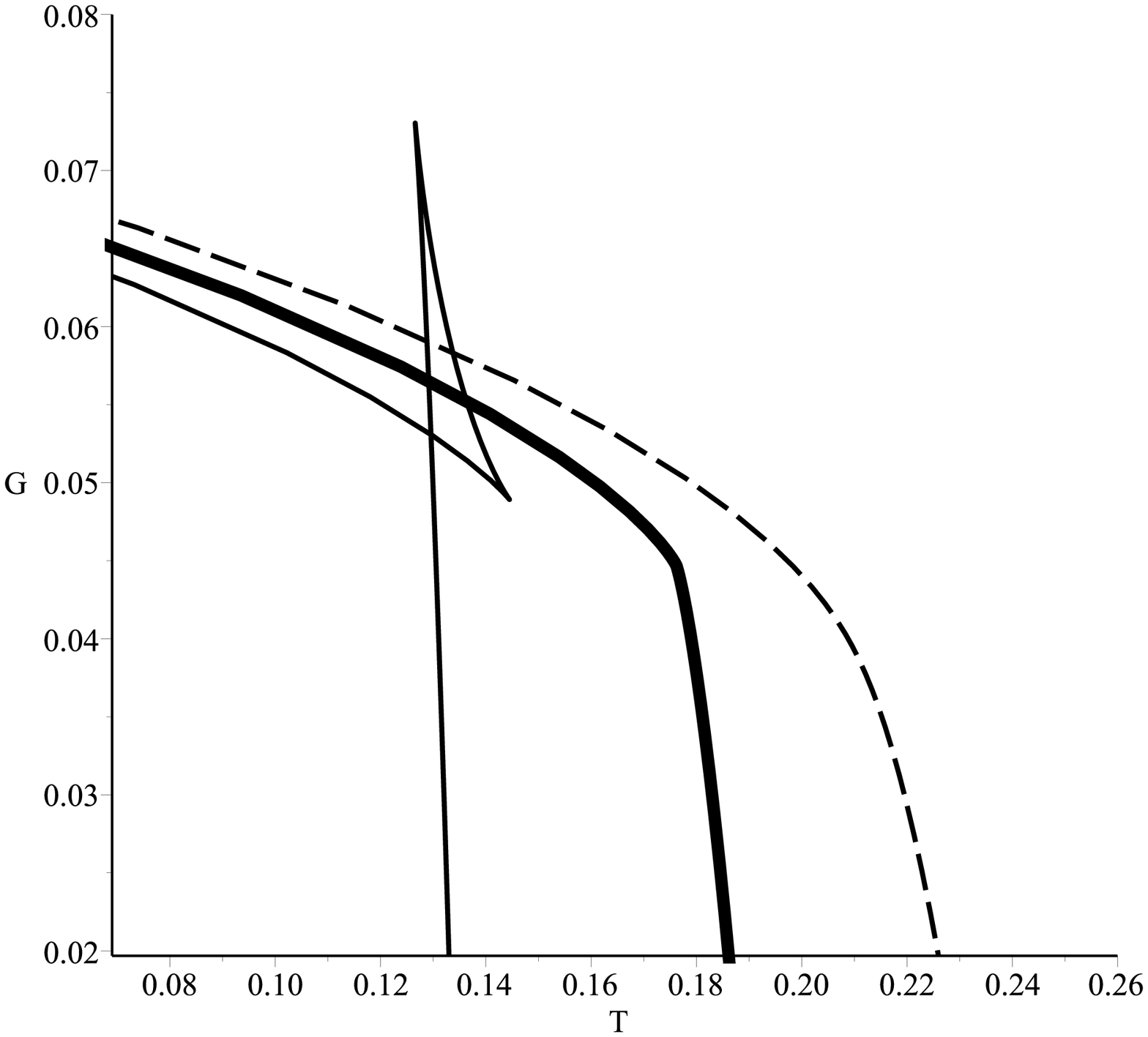}
&
\end{array}
$%
\caption{$P-r_{+}$ (left), $G-T$ (right) diagrams for $c=1$, $n=4$, $q=1$ and $%
\protect\omega =5$.\newline
$P-r_{+}$ diagram, from up to bottom $T=1.8T_{c}$, $T=1.5T_{c}$, $T=T_{c}$, $%
T=0.8T_{c}$ and $T=0.5T_{c}$ respectively.\newline $G-T$ diagram,
from up to bottom $P=1.5P_{c}$, $P=P_{c}$ and $P=0.5P_{c}$,
respectively.} \label{Figw5n4}
\end{figure}
\begin{figure}[tbp]
$%
\begin{array}{ccc}
\epsfxsize=6cm \epsffile{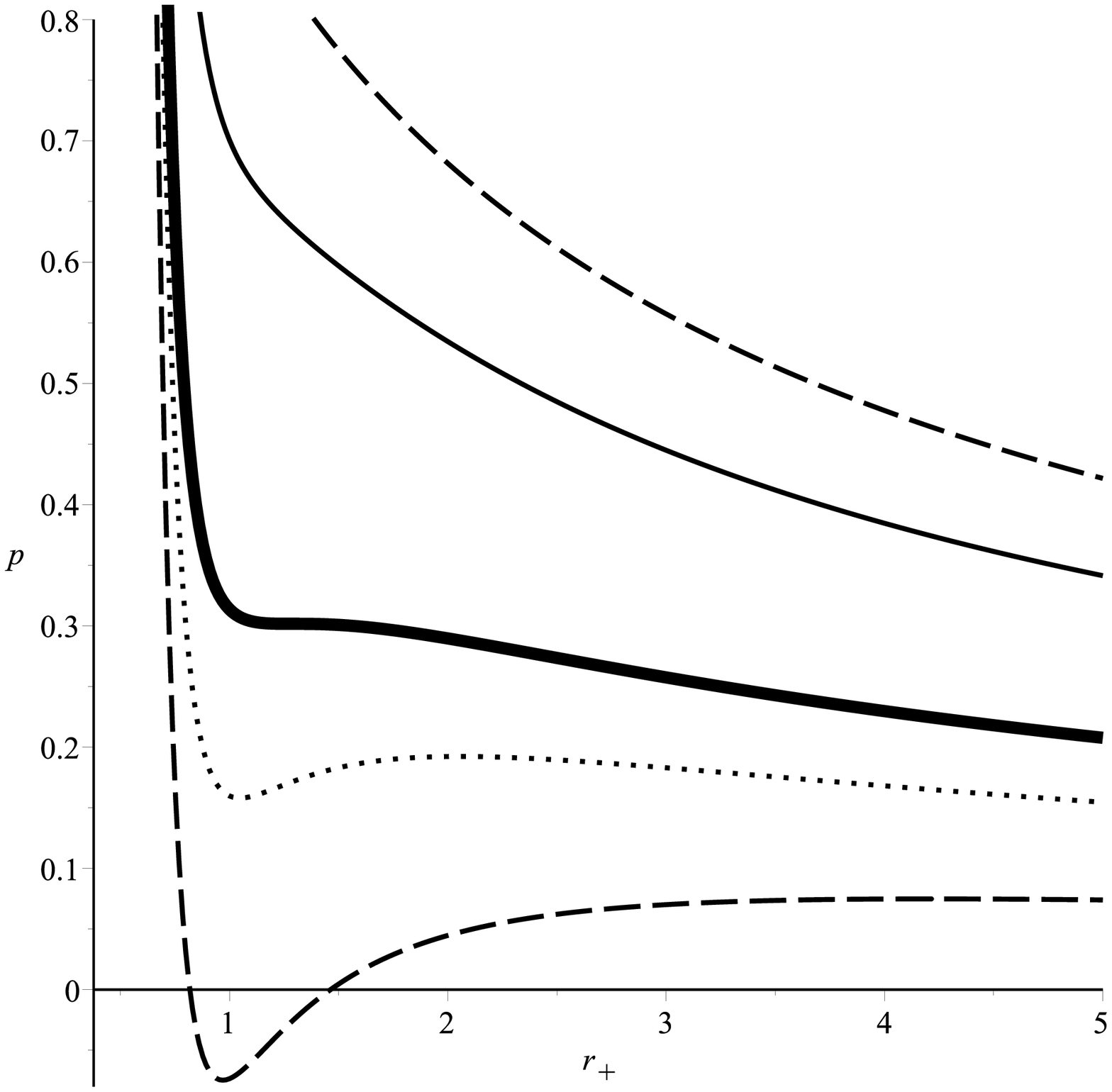} & \epsfxsize=6cm \epsffile{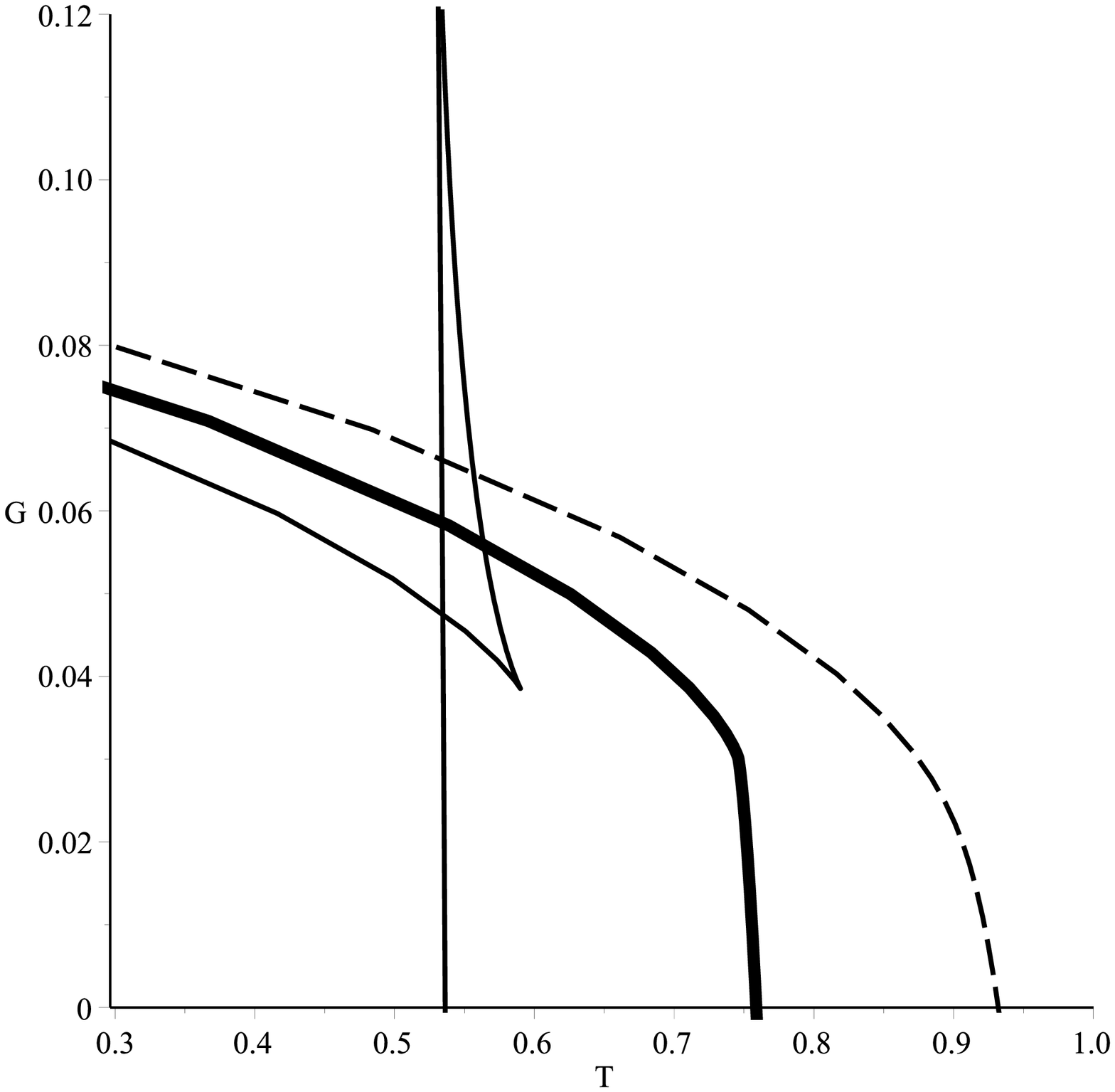}
&
\end{array}
$%
\caption{$P-r_{+}$ (left), $G-T$ (right) diagrams for $c=1$, $n=6$, $q=1$ and $%
\protect\omega =1$.\newline
$P-r_{+}$ diagram, from up to bottom $T=1.8T_{c}$, $T=1.5T_{c}$, $T=T_{c}$, $%
T=0.8T_{c}$ and $T=0.5T_{c}$ respectively.\newline $G-T$ diagram,
from up to bottom $P=1.5P_{c}$, $P=P_{c}$ and $P=0.5P_{c}$,
respectively.} \label{Figw1n6}
\end{figure}
\begin{figure}[tbp]
$%
\begin{array}{ccc}
\epsfxsize=6cm \epsffile{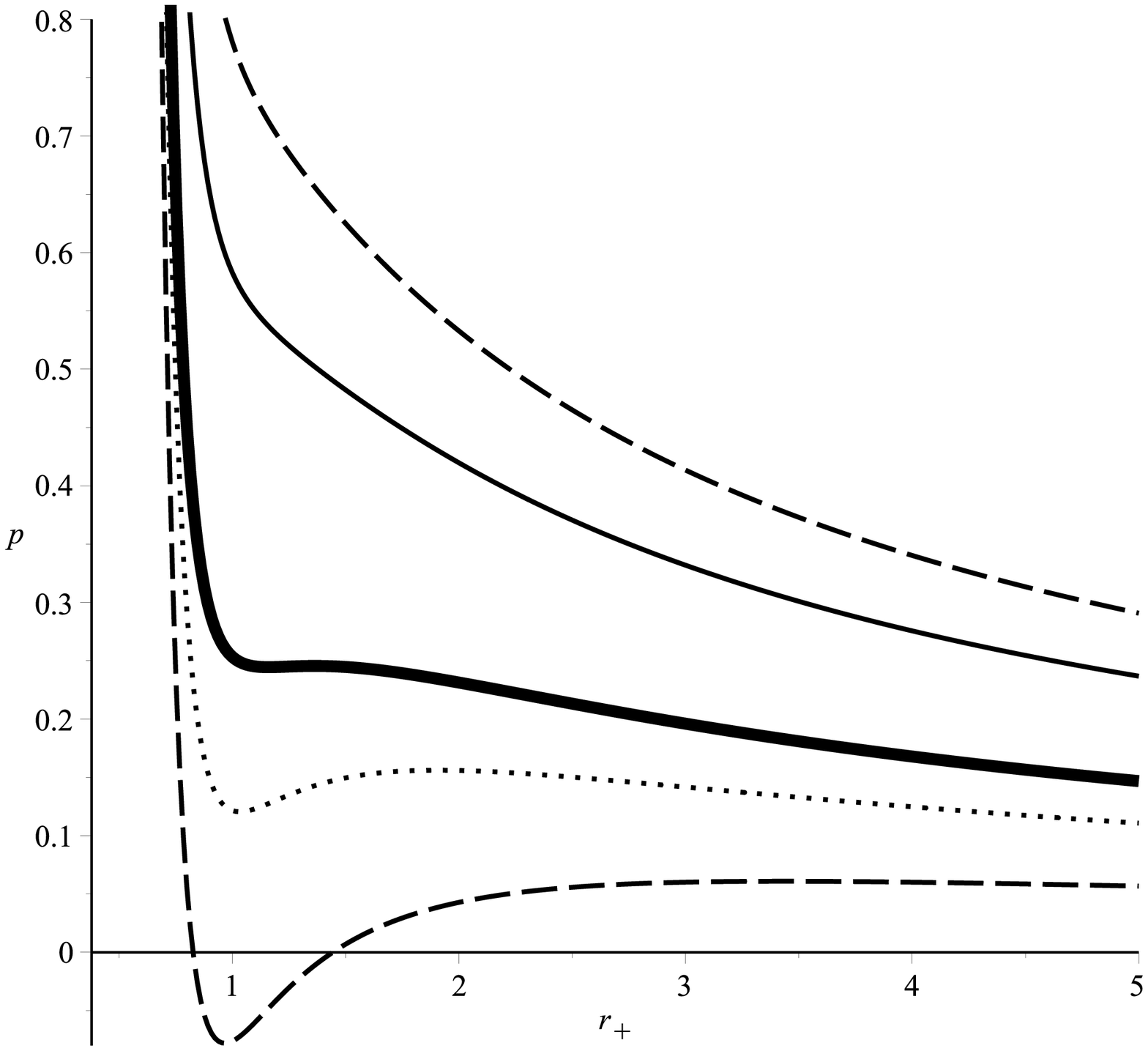} & \epsfxsize=6cm \epsffile{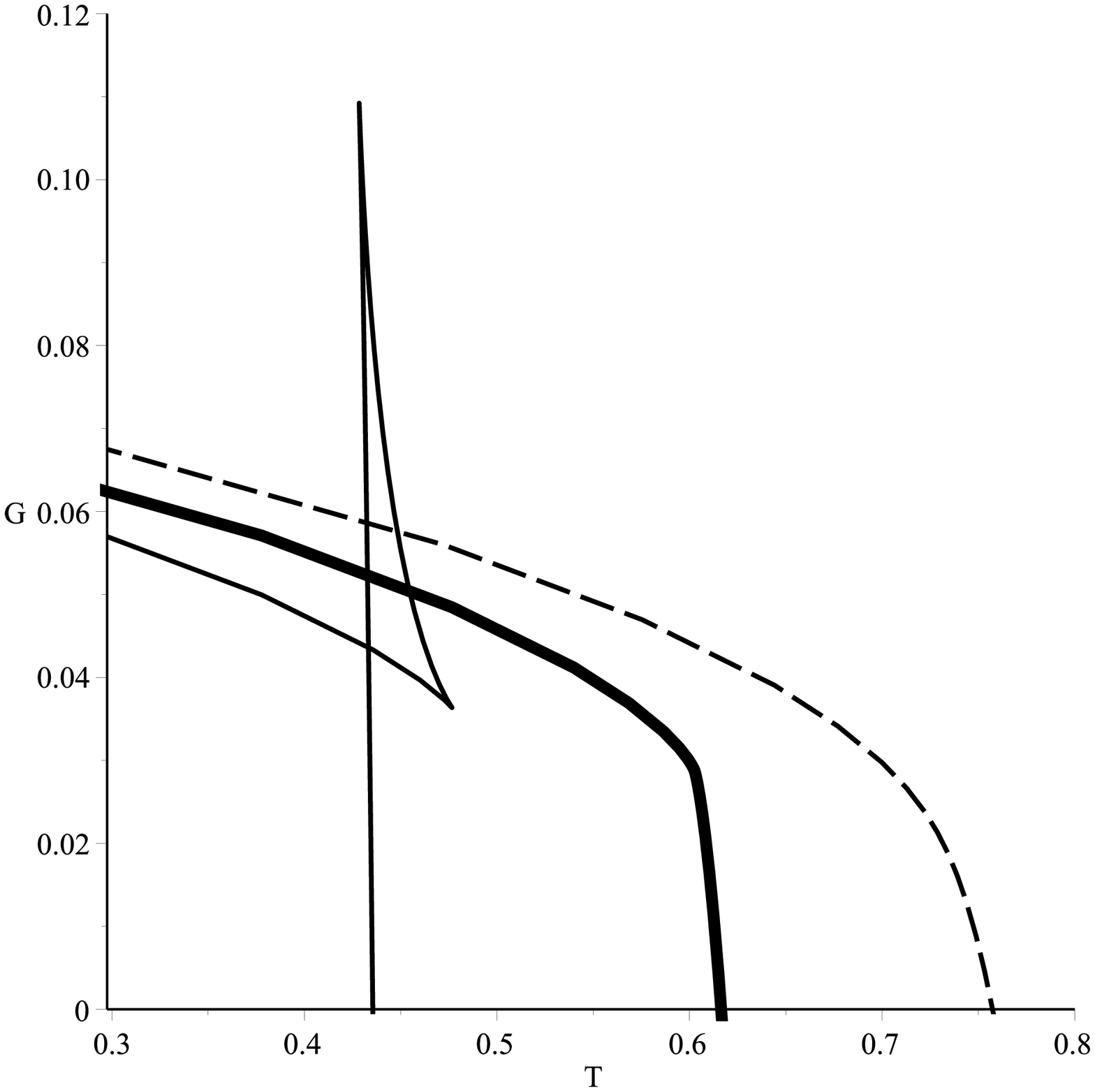}
&
\end{array}
$%
\caption{$P-r_{+}$ (left), $G-T$ (right) diagrams for $c=1$, $n=6$, $q=1$ and $%
\protect\omega =3$.\newline
$P-r_{+}$ diagram, from up to bottom $T=1.8T_{c}$, $T=1.5T_{c}$, $T=T_{c}$, $%
T=0.8T_{c}$ and $T=0.5T_{c}$ respectively.\newline $G-T$ diagram,
from up to bottom $P=1.5P_{c}$, $P=P_{c}$ and $P=0.5P_{c}$,
respectively.} \label{Figw3n6}
\end{figure}
\begin{figure}[tbp]
$%
\begin{array}{ccc}
\epsfxsize=6cm \epsffile{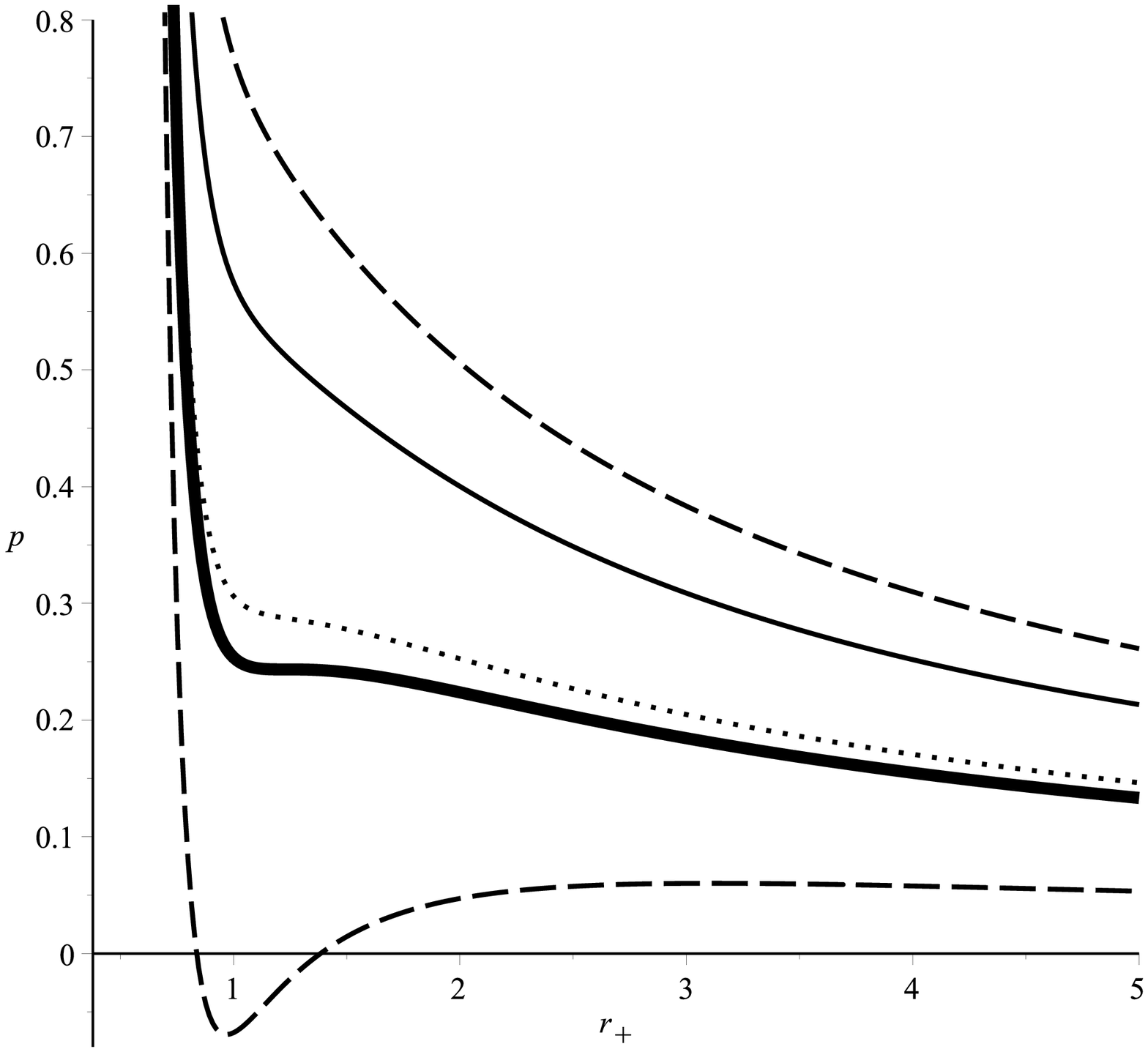} & \epsfxsize=6cm \epsffile{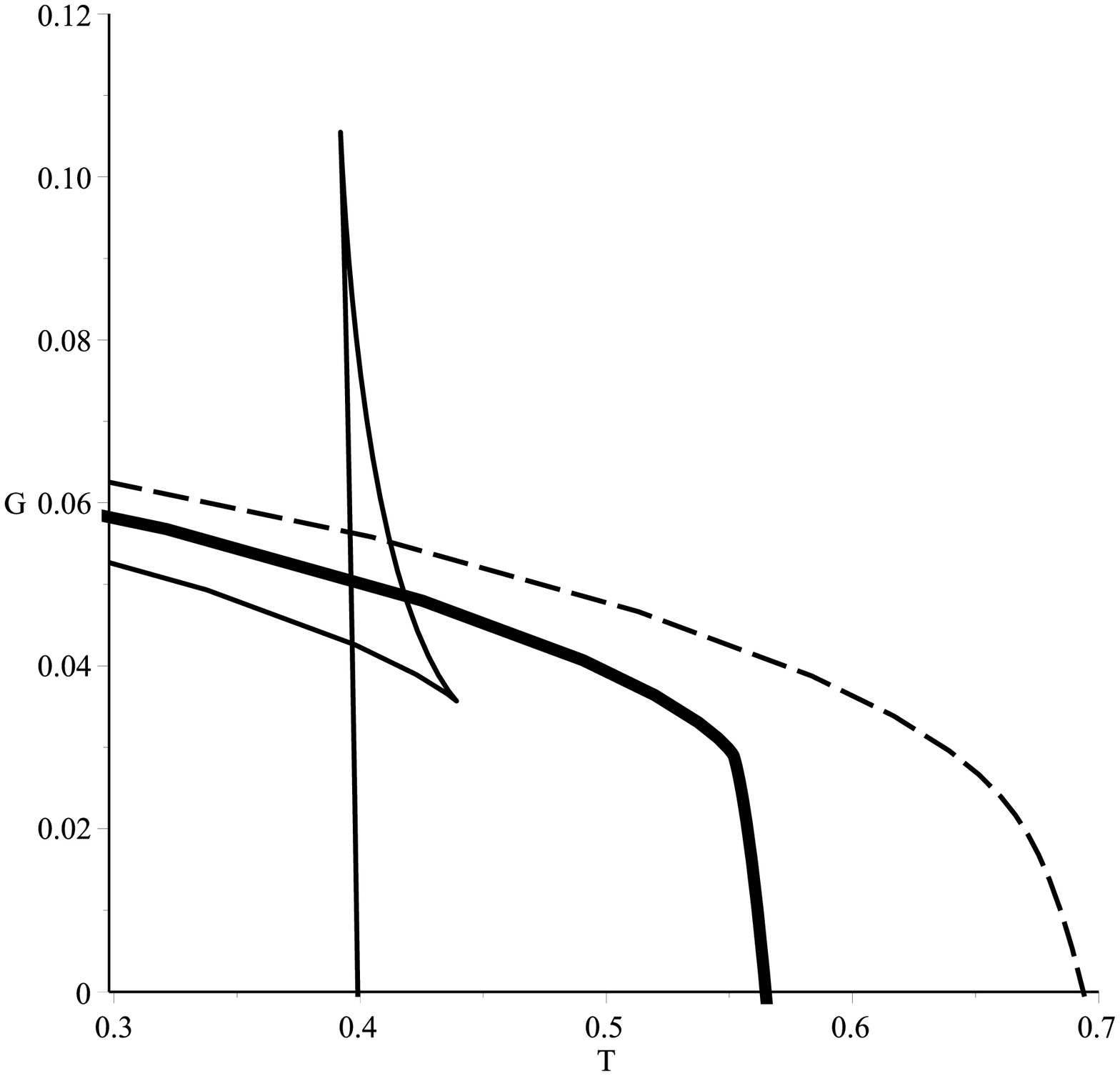}
&
\end{array}
$%
\caption{$P-r_{+}$ (left), $G-T$ (right) diagrams for $c=1$, $n=6$, $q=1$ and $%
\protect\omega =5$.\newline
$P-r_{+}$ diagram, from up to bottom $T=1.8T_{c}$, $T=1.5T_{c}$, $T=T_{c}$, $%
T=0.8T_{c}$ and $T=0.5T_{c}$ respectively.\newline $G-T$ diagram,
from up to bottom $P=1.5P_{c}$, $P=P_{c}$ and $P=0.5P_{c}$,
respectively.} \label{Figw5n6}
\end{figure}
\begin{figure}[tbp]
$%
\begin{array}{ccc}
\epsfxsize=6cm \epsffile{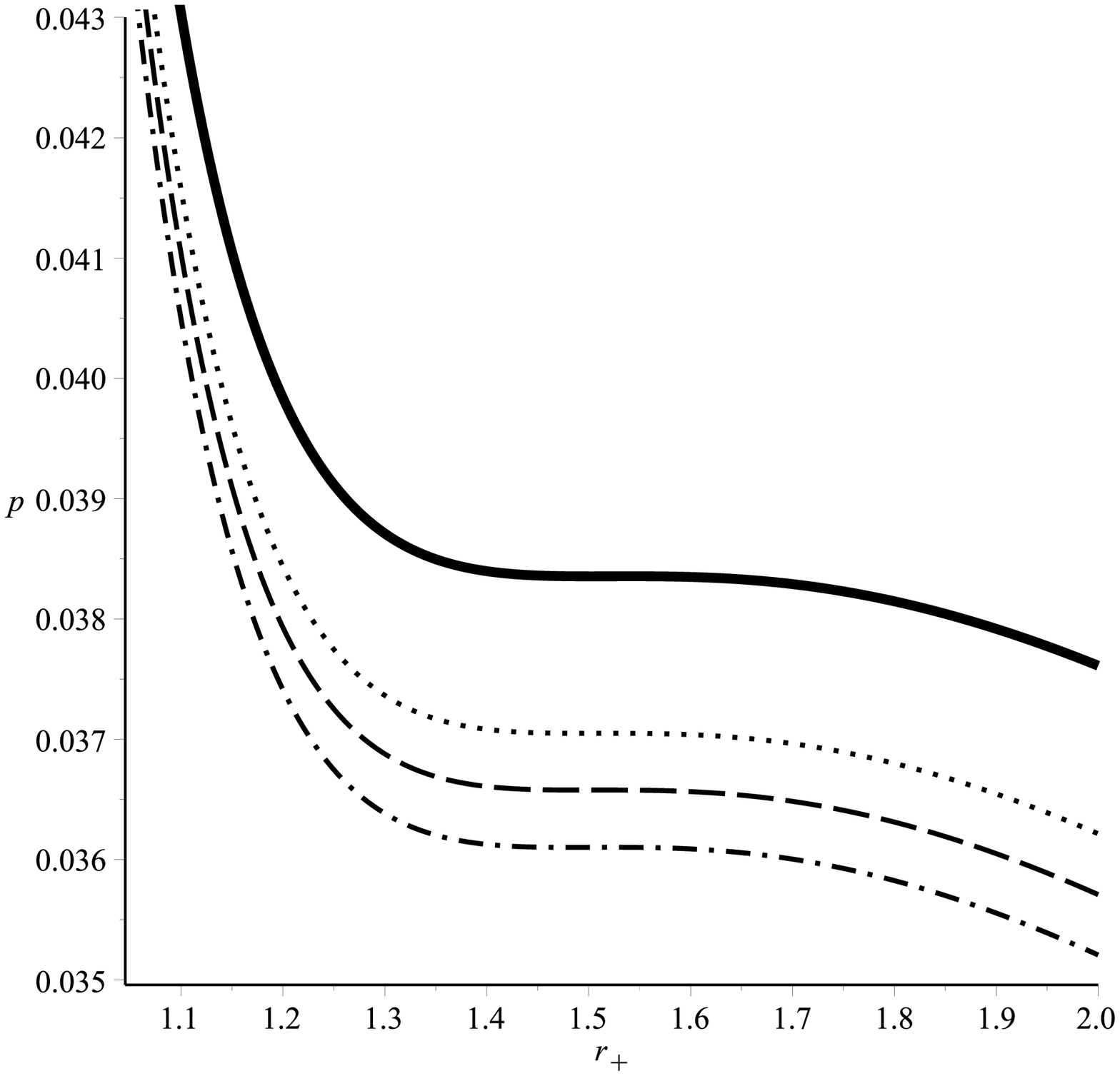} & \epsfxsize=6cm \epsffile{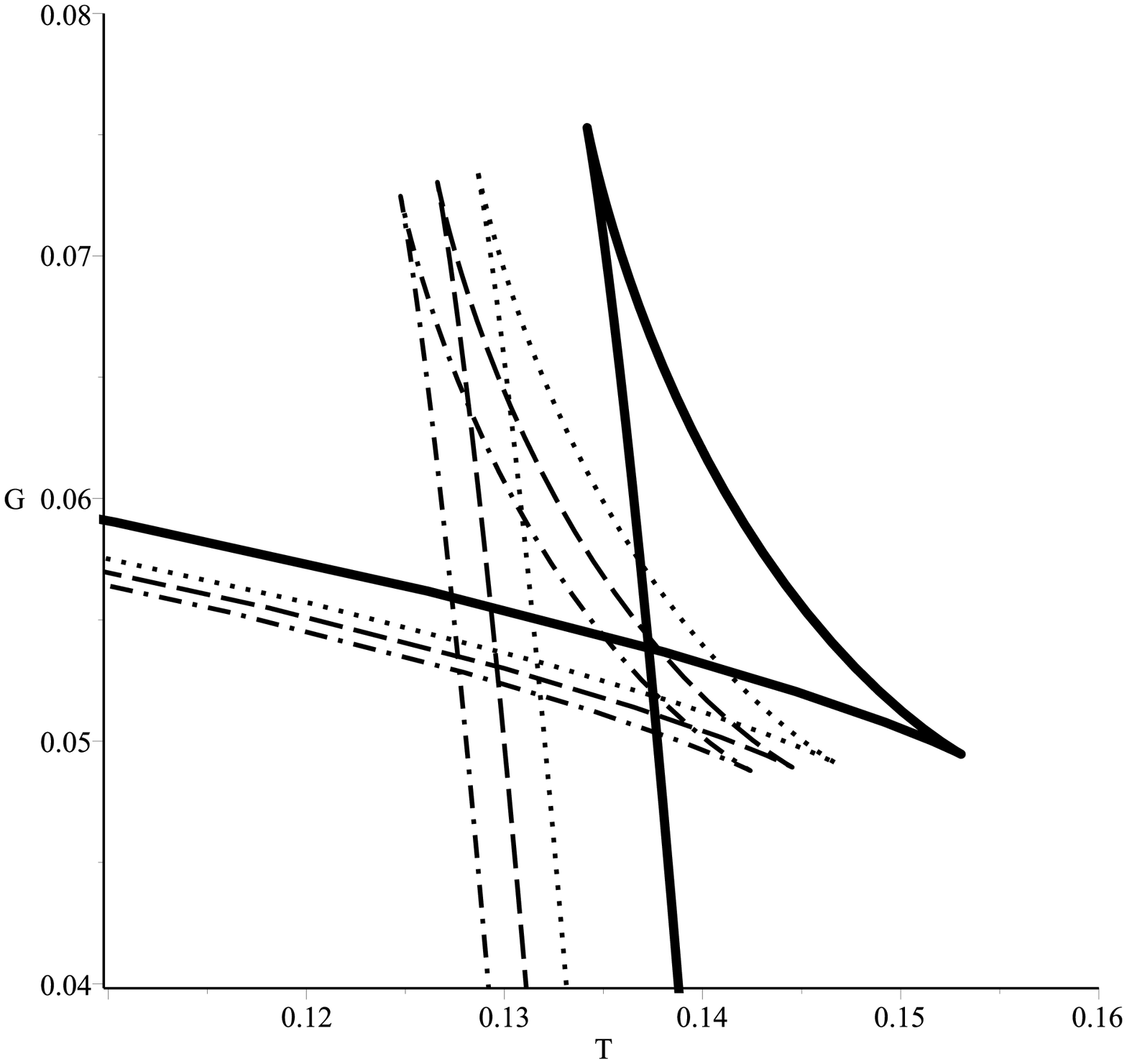} &
\end{array}
$%
\caption{$P-r_{+}$ (left), $G-T$ (right) diagrams for $c=1$,
$n=4$, $q=1$.\newline
$P-r_{+}$ diagram, for $T=T_{c}$, $\protect\omega =1$ (solid line), $\protect%
\omega =3$ (dotted line), $\protect\omega =5$ (dashed line) and $\protect\omega %
=10$ (dasheddotted line).\newline
$G-T$ diagram, for $P=0.5P_{c}$, $\protect\omega =1$ (solid line), $\protect%
\omega =3$ (dotted line), $\protect\omega =5$ (dashed line) and $\protect\omega %
=10$ (dasheddotted line).} \label{Fign4}
\end{figure}
\begin{figure}[tbp]
$%
\begin{array}{ccc}
\epsfxsize=6cm \epsffile{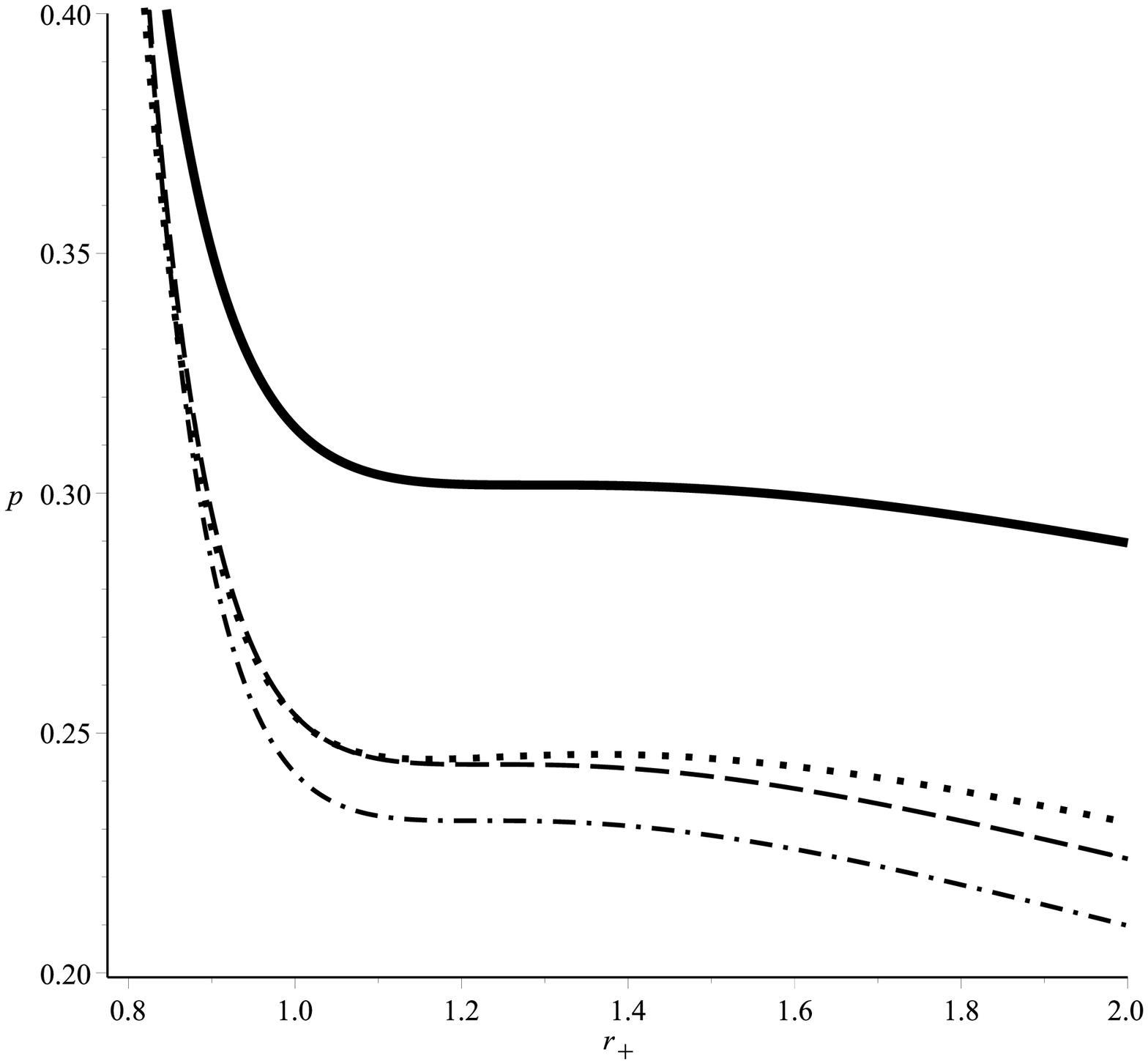} & \epsfxsize=6cm \epsffile{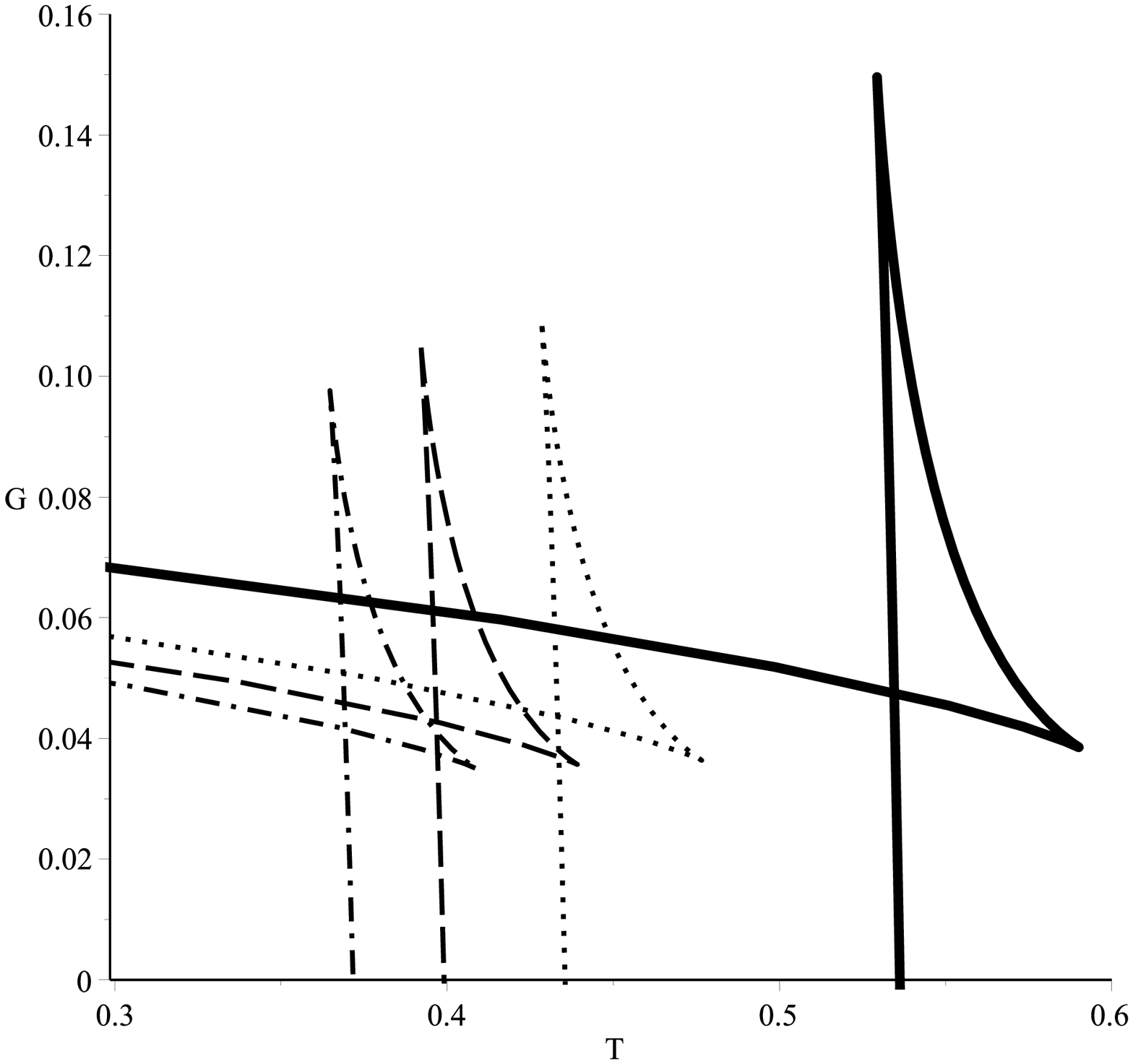} &
\end{array}
$%
\caption{$P-r_{+}$ (left), $G-T$ (right) diagrams for $c=1$,
$n=6$, $q=1$.\newline
$P-r_{+}$ diagram, for $T=T_{c}$, $\protect\omega =1$ (solid line), $\protect%
\omega =3$ (dotted line), $\protect\omega =5$ (dashed line) and $\protect\omega %
=10$ (dasheddotted line).\newline
$G-T$ diagram, for $P=0.5P_{c}$, $\protect\omega =1$ (solid line), $\protect%
\omega =3$ (dotted line), $\protect\omega =5$ (dashed line) and $\protect\omega %
=10$ (dasheddotted line).} \label{Fign6}
\end{figure}
\begin{figure}[tbp]
$%
\begin{array}{ccc}
\epsfxsize=6cm \epsffile{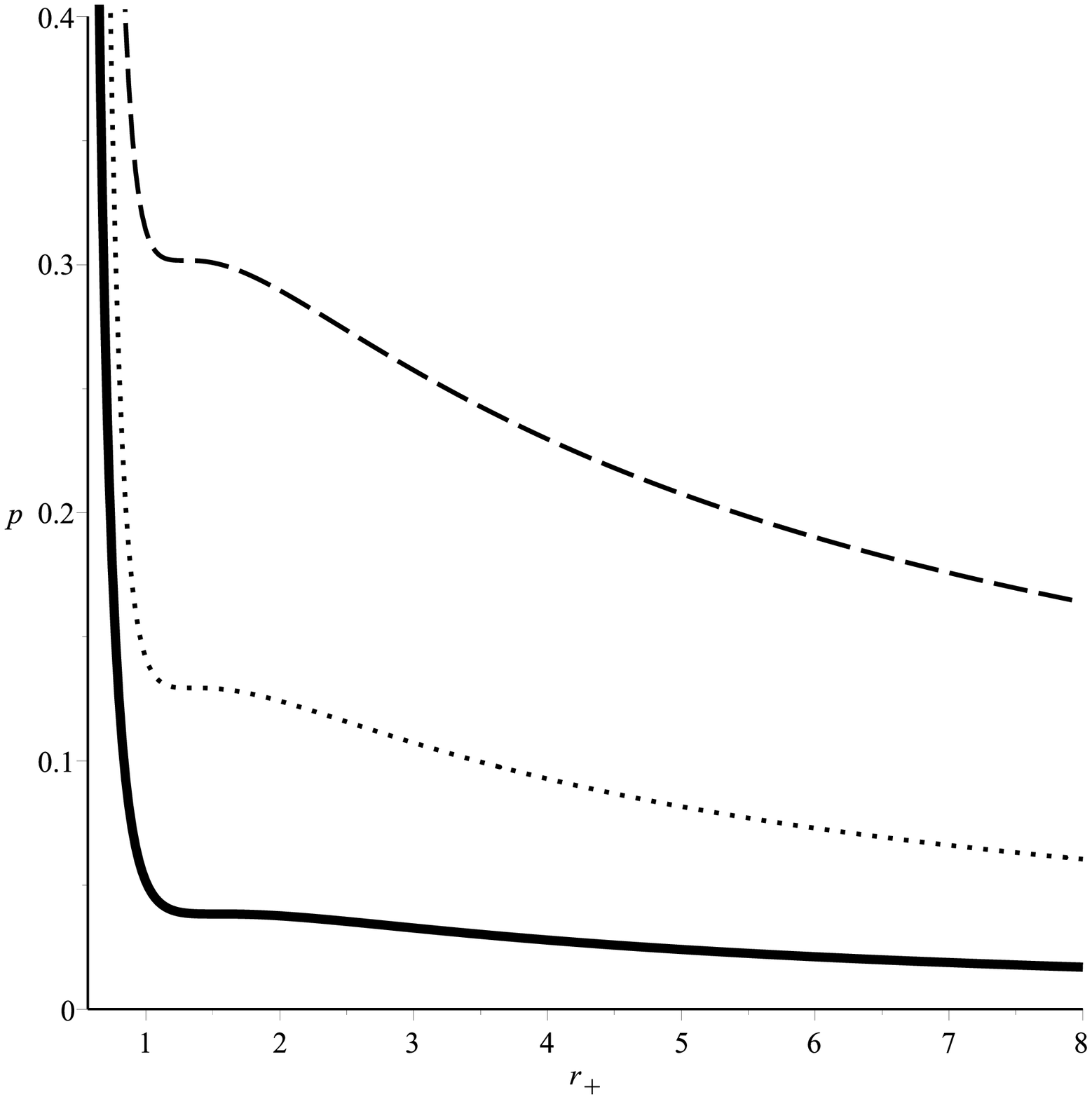} & \epsfxsize=6cm \epsffile{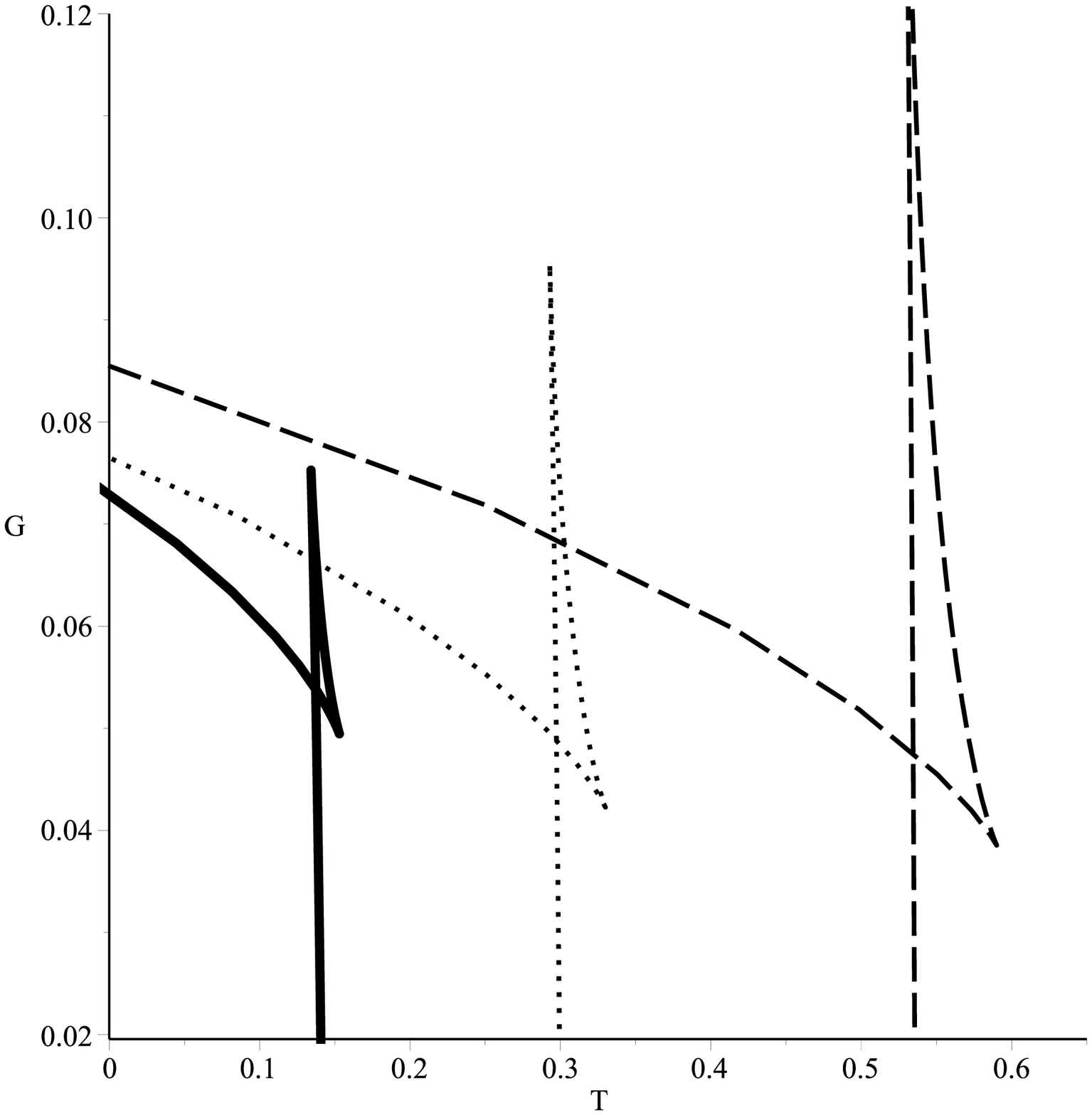} &
\end{array}
$%
\caption{$P-r_{+}$ (left), $G-T$ (right) diagrams for $c=1$,
$\protect\omega =1$, $q=1$.\newline
$P-r_{+}$ diagram, for $T=T_{c}$, $n=4$ (solid line), $n=5$ (dotted line) and $%
n=6$ (dashed line).\newline
$G-T$ diagram, for $P=0.5P_{c}$, $n=4$ (solid line), $n=5$ (dotted line) and $%
n=6$ (dashed line).} \label{Figw1}
\end{figure}

\section{ Discussion on the results of diagrams\label{results}}

Thermodynamical behavior of the system is shown in Figs.
\ref{Figw1n4}-\ref{Figw1}. These critical values represent phase
transition points, which one can see in $G-T$ and $P-r_{+}$
diagrams. Studying $P-r_{+}$ graphs (left panel of Figs.
\ref{Figw1n4}-\ref{Figw1}) show that obtained values are critical
points in which phase transition occurs. It can be seen in the
graphs that as the coupling constant ($\omega$) increases, the
temperature in which phase transition occurs decreases. Increasing
$\omega$ leads to decreasing in the value of Gibbs free energy of
phase transition point (see Fig. \ref{Fign4} for more details).
The results indicate that with larger coupling constant the energy
that system needs in order to have phase transition becomes less.
It is also evident from the effect of the coupling constant on the
total finite mass of the black hole that by increasing the
coupling constant the total finite mass which is interpreted as
enthalpy of the system will increase too. It means that in order
to have phase transition it is expected that the mentioned black
hole absorb more mass from surrounding.

Studying $P-r_{+}$ diagrams shows that as coupling constant
increases, both critical pressure and horizon radius of critical
point decrease. On the other hand, it is worthwhile to mention
that due to the relation between pressure and cosmological
constant (which is related to the asymptotical curvature of the
background), as the coupling constant increases, the necessity of
having a background with more curvature increases too.

Using Fig. \ref{Figw1}, we can also discuss the effect of
dimensionality on critical point and behavior of system. It can be
seen from the obtained $G-T$ diagrams for different dimensions
that as dimensionality increases, the temperature of critical
point increases too. According to $P-r_{+}$ diagrams the pressure
in which phase transition occurs will increases as dimensionality
increases. As a consequence of increment in pressure one can see
that the cosmological constant decreases, due to the relation of
$P=-\frac{\Lambda }{8\pi }$. Hence the need of having a background
with higher value of curvature decreases in higher dimensions. It
can be inferred that as dimensionality increases the system needs
to absorb more mass in order to have phase transition.

At the end one can see the critical horizon radius decreases as
coupling constant increases same as critical temperature and
pressure. On the other hand for higher dimensions, we have higher
values of critical points. The results arisen from graphs can also
be seen directly through the tables I and II.


\begin{center}
\begin{tabular}{c}
\begin{tabular}{cccc}
\hline\hline
$\omega$ \;& $r_{c}$ \;& $T_{c}$ \;& $P_{c}$  \\
\hline\hline $1.0000$ \;& $1.5227$ \;& $0.1869$ \;& $0.0384$  \\
\hline $3.0000$ \;& $1.5099$ \;& $0.1790$ \;& $0.0371$  \\
\hline $5.0000$ \;& $1.5052$ \;& $0.1762$ \;& $0.0366$  \\
\hline $10.0000$ \;& $1.5010$ \;& $0.1736$ \;& $0.0361$  \\ \hline
\end{tabular}
\\
Table I: critical quantities for $q=1$ and $n=4$.
\end{tabular}
\end{center}

\begin{center}
\begin{tabular}{c}
\begin{tabular}{cccc}
\hline\hline
$\omega $ \;& $r_{c}$ \;& $T_{c}$ \;& $P_{c}$  \\
\hline\hline $1.0000$ \;& $1.2859$ \;& $0.7460$ \;& $0.3018$  \\
\hline $3.0000$ \;& $1.2456$ \;& $0.5839$ \;& $0.2627$  \\
\hline $5.0000$ \;& $1.2325$ \;& $0.5517$ \;& $0.2434$  \\
\hline $10.0000$ \;& $1.2210$ \;& $0.5130$ \;& $0.2318$  \\ \hline
\end{tabular}
\\
Table II: critical quantities for $q=1$ and $n=6$.
\end{tabular}
\end{center}

\section{Conclusions}

In this paper, we considered the BD theory in the presence of
electromagnetic field and studied its phase structure. We extended
the phase space by considering cosmological constant as
thermodynamical pressure and its conjugate variable as volume and
regarded the interpretation of total mass of black hole as the
enthalpy of the system.

Studying calculated critical values through two different types of
phase diagrams resulted into phase transition taking place in the
critical values. Studying $P-r_{+}$ and $G-T$ diagrams exhibits
similar behavior near critical points to their corresponding
diagrams in Van der Waals liquid/gas.

The results indicated that for large values of coupling constant
the system needs less energy (mass) absorption to have phase
transition, due to the fact that as coupling constant increases
the critical temperature decreases. On the other hand, studying
the effects of dimensionality showed that for higher dimensional
black holes, phase transition take places in higher temperature
and lower Gibbs free energy.

Finally considering BD theory with various models of nonlinear
electrodynamics, it would be interesting to analyze the effects of
nonlinearity on extended phase space thermodynamics and $P-V$
criticality of black hole solutions. We left these issues for the
forthcoming work.

\acknowledgments{We thank Shahram Panahiyan for reading the
manuscript. We also wish to thank the Shiraz University Research
Council. This work has been supported financially by Research
Institute for Astronomy and Astrophysics of Maragha, Iran.}

\end{document}